\documentclass{l4dc2025}

\title[Safe and Stable Neural ODE Control]{Opt-ODENet: A Neural ODE Framework with Differentiable QP Layers for Safe and Stable Control Design (longer version)}
\usepackage{times}
\usepackage{tikz}
\usepackage{mathtools}
\usepackage{newtxtext, newtxmath}
\allowdisplaybreaks

\coltauthor{%
\Name{Keyan Miao}\textsuperscript{1} \Email{keyan.miao@eng.ox.ac.uk}\\
\Name{Liqun Zhao}\textsuperscript{1} \Email{liqun.zhao@eng.ox.ac.uk}\\
\Name{Han Wang}\textsuperscript{1} \Email{han.wang@eng.ox.ac.uk}\\
\Name{Konstantinos Gatsis}\textsuperscript{2,3} \Email{konstantinos.gatsis@villanova.edu}\\
\Name{Antonis Papachristodoulou}\textsuperscript{1} \Email{antonis@eng.ox.ac.uk}\\
\addr \textsuperscript{1}Department of Engineering, University of Oxford, UK \\
\addr \textsuperscript{2} Department of Electrical and Computer Engineering, Villanova University, USA \\
\addr \textsuperscript{3} School of Electronics and Computer Science, University of Southampton, UK
}

\begin{document}

\maketitle

\begin{abstract}%
Designing controllers that achieve task objectives while ensuring safety is a key challenge in control systems. This work introduces Opt-ODENet, a Neural ODE framework with a differentiable Quadratic Programming (QP) optimization layer to enforce constraints as hard requirements. Eliminating the reliance on nominal controllers or large datasets, our framework solves the optimal control problem directly using Neural ODEs. Stability and convergence are ensured through Control Lyapunov Functions (CLFs) in the loss function, while Control Barrier Functions (CBFs) embedded in the QP layer enforce real-time safety. By integrating the differentiable QP layer with Neural ODEs, we demonstrate compatibility with the adjoint method for gradient computation, enabling the learning of the CBF class-$\mathcal{K}$ function and control network parameters. Experiments validate its effectiveness in balancing safety and performance.
\end{abstract}


\section{Introduction}
Designing safe and efficient control systems is crucial in automation and control. Controllers must accomplish tasks while ensuring safety in environments under strict constraints. While learning-based control has shown promise in complementing traditional methods, integrating safety guarantees with task performance remains challenging. In this work, we propose a learning-based control framework that combines Neural Ordinary Differential Equations (Neural ODEs) with Control Barrier Functions (CBFs) and Control Lyapunov Functions (CLFs) to address this challenge.\\
Neural ODEs \citep{NODE}, provide a natural framework for modeling and optimizing continuous-time dynamic behaviors by treating networks as systems governed by differential equations. They offer adaptability, flexibility, and smooth gradient computation, making them well-suited for applications such as trajectory planning, state observer design, and system identification \citep{liang2021modeling, miao2023learning, djeumou2022neural,bottcher2022ai}. However, ensuring task completion and safety in real-world applications requires additional mechanisms.\\
CBFs \citep{ames2019control, ames2016control} and CLFs \citep{ames2014rapidly} are widely used in control theory to enforce safety constraints and ensure stability by defining safe sets and guaranteeing convergence to target states, respectively. Recently, CBFs and CLFs have been incorporated into learning-based frameworks, where they are often used as soft constraints to guide control design \citep{jin2020neural, zhao2023stable, zhao2025nlbac} or to improve performance in machine learning tasks such as image classification \citep{pmlr-v235-miao24a}. Furthermore, the co-design of controllers with these functions has enabled adaptive and robust control in dynamic environments \citep{wang2023safety, wang2024convex, dawson2022safe}.\\
To enforce constraints as hard requirements, we incorporate Quadratic Programming (QP) into a learning framework with differentiable optimization layer \citep{amos2017optnet}. QP embeds CBFs to enforce real-time safety constraints, while CLFs are incorporated into the loss function to ensure stability and task performance without conflicting with hard constraints. Its differentiability enables gradient-based optimization, allowing learning parameters of class $\mathcal{K}$ function for CBF constraints and preceding network parameters.\\
Several works closely related to learning-based control with safety considerations have been proposed, yet they exhibit certain limitations. Some rely on soft constraints to handle safety, which may fail in critical scenarios or only consider stability properties \citep{sandoval2023neural, ip2024lyapunov}. Others utilize fixed CBF parameters, limiting their adaptability to varying environments \citep{pereira2021safe}. Certain approaches depend on a nominal controller to guide control \citep{taylor2020learning}, with some adopting supervised learning that relies on nominal safe controllers or reference trajectories and require large sampled datasets \citep{xiao2023barriernet, xiao2023safe, bachhuber2023neural}. Furthermore, some works focus primarily on motion planning, adjusting abstract signals to satisfy constraints rather than directly controlling physical input signals \citep{10611584}.\\
The main contributions of this work are as follows: 1) We propose Opt-ODENet, a Neural ODE framework integrating differentiable QP layers for safe and stable control, eliminating the need for a nominal controller and reducing reliance on pre-designed policies or large datasets. 2) We incorporate CLFs into the loss function for stability and convergence, and embed CBFs as hard constraints into a differentiable QP layer, demonstrating compatibility with Neural ODEs and gradient computation via the adjoint method. 3) The framework adaptively learns the class-$\mathcal{K}$ function for CBF, enabling a flexible balance between safety and task performance. 4) Through experiments, we validate the proposed framework and show the applicability of higher-order CBFs (HOCBFs) within our framework, extending its capabilities to more complex systems.
\section{Preliminaries}
\subsection{Neural ODEs with input}
Residual Networks (ResNet) \citep{ResNet}, a milestone in deep learning, introduced skip connections that inspired viewing layers as discrete-time dynamical systems. Extending this idea, Neural ODEs \citep{NODE} model dynamics in continuous time, described as:
\begin{definition}[Neural ODE with input]
A Neural ODE with input is a system of the form
\begin{equation}
\left\{
\begin{array}{ll}
\dot {x} \left(t\right) = \mathcal{F}\left(t, x\left(t\right), u\left(t\right), \theta\right), \quad t \in \mathcal{S} & \\
x\left(t_0\right) = x_0
\end{array}
\right.
\label{node}
\end{equation}
where $\mathcal{S}:= \left[t_0, t_f\right]$ $(t_0, t_f\in \mathbb{R}^+)$ is the depth domain and $\mathcal{F}$ is a neural network referred to as ODENet with parameter $\theta$; $u(t)$ is the input at time $t$.
\end{definition}
The terminal state $x(t_f)$, obtained by solving the initial value problem (IVP), represents the evolved system state. In Neural ODEs, depth corresponds to continuous progression along the time domain, with ResNet interpreted as a discretization using the Euler method.\\
In applications like image classification, Neural ODEs map an initial state $x(t_0)$ to a terminal state $x(t_f)$, e.g., a class label, by solving the initial value problem:
$x(t_f) = x(t_0) + \int_{t_0}^{t_f} \mathcal{F}(t, x(t), u(t), \theta)\,dt$. Training typically involves minimizing a loss that depends only on the terminal state $x(t_f)$, such as cross-entropy, which corresponds to a Mayer-type optimal control problem. More generally, a loss can be defined along the entire trajectory: $\ell := \Phi(x(t_f)) + \int_{t_0}^{t_f} \mathcal{L}(x(t), u(t), \theta)\,dt$, where $\Phi$ is a terminal cost and $\mathcal{L}$ is a running cost. This defines a Bolza-type problem \citep{MP}, which generalizes Mayer by including path-dependent penalties (e.g., regularization). Training then corresponds to solving this Bolza objective via gradient-based optimization.
\begin{equation}
\begin{aligned}
& \min_{\theta \in U} \quad \ell \\
s.t. & \quad \dot{x}(t) = \mathcal{F}(t, x(t), u(t), \theta), \, t \in \mathcal{S}, \\
& \quad x(t_0) = x_0,
\end{aligned}
\end{equation}
\subsection{CLF and CBF}
Consider the dynamical system
\begin{equation}\label{eq:system}
    \dot x(t) = \mathcal{F}(x(t),u(t)),
\end{equation}
where $\mathcal{F}(x(t),u(t)):\mathcal{X}\times \mathcal{U}\to \mathbb{R}^{d_x}$ is a locally Lipschitz continuous function and $\mathcal{X} \subseteq \mathbb{R}^{d_x}$ and $\mathcal{U} \subseteq \mathbb{R}^{d_u}$ are compact sets. The controller $u(t)$ is designed for two objectives, namely ensuring \emph{stability} and \emph{safety}. The concept of safety involves an additional set $\mathcal{B}\subseteq \mathcal{X}$ which interprets certain physical requirements, such as collision avoidance for robots, lane keeping for vehicles, etc. Mathematically speaking, we say dynamical system \eqref{eq:system} is \emph{safe} if there exists a locally Lipschitz continuous controller $\tau(x(t)):\mathcal{X}\to\mathcal{U}$ with $u(t)=\tau(x(t))$, such that $x(t)\in\mathcal{B}$ for any $x(0)\in\mathcal{B}$ and $t\ge 0$. One prevalent way to design a controller that ensures stability and safety relies on certificate functions. For stability, a Control Lyapunov Function (CLF) is defined as follows.

\begin{definition}[Control Lyapunov Function \citep{sontag1989universal}]
Consider system \eqref{eq:system}. A locally positive definite and differentiable function $V(\cdot):\mathcal{X}\to\mathbb{R}$ is called a Control Lyapunov Function if there exists a locally Lipschitz continuous controller $\tau(x(t)):\mathcal{X}\to\mathcal{U}$, such that
\begin{equation}\label{eq:clf}
    \frac{\partial V(x)}{\partial x}\mathcal{F}(x,\tau(x))+\gamma(V(x))\le0,\forall x\in\mathcal{X},
\end{equation}
where the function $\gamma(\cdot):\mathbb{R}^+\to\mathbb{R}$ is of class-$\mathcal{K}$, i.e. $\gamma(0)=0$ and $\gamma(\cdot)$ is strictly increasing.
\end{definition}
For safety, a Control Barrier Function (CBF) is defined as follows.
\begin{definition}[Control Barrier Function \citep{ames2019control}]
    Consider system \eqref{eq:system} and the safe set $\mathcal{B}$. A differentiable function $B(x):\mathcal{X}\to\mathbb{R}$ is called a Control Barrier Function (CBF) if $\mathcal{B}:=\{x\in\mathcal{X}:B(x)\ge0\}$, and there exists a locally Lipschitz continuous controller $\tau(x(t)):\mathcal{X}\to\mathcal{U}$, such that
    \begin{equation}\label{eq:cbf}
        \frac{\partial B(x)}{\partial x}\mathcal{F}(x,\tau(x))+\alpha(B(x))\ge 0,\forall x\in\mathcal{B}.
    \end{equation}
    The function $\alpha(\cdot):\mathbb{R}\to \mathbb{R}$ is of extended class-$\mathcal{K}$.
\end{definition}
\subsection{Differentiable Optimization Layer}
A differentiable optimization problem refers to a class of optimization problems whose solutions can be differentiated through backpropagation. This feature allows such optimization problems to function as layers within deep learning architectures, enabling the encoding of constraints and complex dependencies that traditional convolutional and fully connected layers typically cannot capture. In this work, we incorporate the differentiable optimization layer OptNet proposed by \cite{amos2017optnet} to implement the CBF-QP layer to incorporate the safety constraint.\\
OptNet defined a neural network layer based on quadratic programming problem:
\begin{equation}
    \begin{aligned}
        z_{i+1} = & \arg \min_z \frac{1}{2}z^{\top}Q(z_i)z + q(z_i)z^{\top}\\
        s.t. \quad &  A(z_i)z=b(z_i), G(z_i)z\leq h(z_i)
    \end{aligned}
    \label{qp}
\end{equation}
where $Q\in \mathbb R^{n\times n}$, $q\in \mathbb R^n$, $A\in \mathbb 
R^{m\times n}$, $b\in \mathbb R^m$, $G\in \mathbb R^{p\times n}$, $h\in \mathbb R^p$ and $Q\geq0$. $z_i$ is the output of the previous layer, $z_{i+1}$ is the output of the optimization layer.\\
The forward pass of the layer consists of formulating and solving the optimization problem. More critically, the backward pass requires computing the derivatives of the QP solution with respect to its input parameters, which are obtained by differentiating the Karush-Kuhn-Tucker (KKT) conditions.
\section{Problem Formulation}
Consider a control-affine system
\begin{equation}
    \dot x = f(x)+g(x)u
    \label{eq: control-affine system}
\end{equation}
where $x$ is the state, $u$ is the control input, and the locally Lipschitz continuous functions $f(x)$, $g(x)$ are known. Furthermore, we assume a CBF $B(x)$ that defines the safe set $\mathcal{B}:=\{x:B(x)\ge 0\}$ is known \emph{a priori}. This is a reasonable assumption since $f(x)$ and $g(x)$ are both known and time-invariant. Our goal is to design a state-feedback control policy $u(t) = \tau(x(t),\theta_1,\theta_2)$, parameterized by $\theta_1$ and $\theta_2$, that satisfies the following requirements:
\vspace{-0.2cm}
\begin{itemize}
\itemsep-0.2em
    \item [(1)] Task achievement: Stabilizes the system \eqref{eq: control-affine system} around a desired terminal state $\hat x$ from an arbitrary initial point $x_0$.
    \item [(2)] Constraint satisfaction: Ensures safety (CBF constraints) of the closed-loop system $\dot x=f(x)+g(x)\tau(x, \theta_1,\theta_2)$. 
    \item [(3)] Performance and Safety trade-off: Optimizes task performance while tuning the strength of the CBF safety constraint via function $\alpha(\theta_2)$, enabling trade-offs along the trajectory.
\end{itemize}
\vspace{-0.2cm}
By achieving a minimum safety margin, we mean that the system should avoid aggressively approaching the boundary of the safe set $\mathcal{B}$ to optimize the control performance, such as minimizing stabilization time.\\
Motivated by the requirements, we propose the following safety-constrained optimal control problem:
\begin{eqnarray}
 \min_{\theta=(\theta_1,\theta_2) \in U} && \ell = \Phi\left(x \left(t_f\right), \hat x\right) + \int_{t_0}^{t_f}\mathcal{L}\left(x\left(t\right), u\left(t\right), \theta\right)dt \notag\\
\text{s.t.} && \dot {x} \left(t\right) = \mathcal{F}(x, u) = f(x)+g(x)u, \quad t \in \mathcal{S} \notag \\
&& \dot B(x(t))+\alpha(\theta_2) B(x(t)) \geq 0,\quad t\in\mathcal{S}
\label{eq:formulation}\\
&& x\left(t_0\right) = x_0, \quad u(t) = \tau(x, \theta_1,\theta_2) \notag
\end{eqnarray}
Here, $\theta = (\theta_1, \theta_2)$ represents the parameters of the problem, with $\theta_1$ parameterizing the control network and $\theta_2$ characterizing the class-$\mathcal{K}$ function $\alpha(\cdot)$ for the CBF constraints.
\section{Proposed Approach}
Solving problem \eqref{eq:formulation} is challenging not only because of the optimal control nature, but also due to the hard safety constraints. To address these difficulties, we propose a learning-based control framework that effectively leverages Neural ODEs and differentiable optimization layers. The framework consists of the following components:
\begin{itemize}
\item Neural ODE Dynamics: The system dynamics and the control policy are parameterized using a Neural ODE $\dot x = \mathcal{F}(x,\tau(x,\theta_1, \theta_2))$ to allow for continuous-time modeling.
\item Differentiable CBF-QP Layer: Safety constraints are enforced through a CBF, embedded within a differentiable QP layer.
\begin{eqnarray}
        u_{safe} = & \arg \min_u & \frac{1}{2}u^{\top}Q(u_{nn})u + q(u_{nn})^{\top}u \notag\\
        &\text{s.t.} &  Au=b, Gu\leq h
    \label{eq:qp-u}
\end{eqnarray}
where in our problem, $Q = I$, $q = -u_{nn}$, $A$ and $b$ are absent, $G = -\frac{\partial B(x)}{\partial x}g(x)$ and $h = \frac{\partial B(x)}{\partial x}f(x)+\alpha(B(x), \theta_2)$. $u_{nn} = \pi(x,\theta_1)$ where $\pi$ represents the controller neural network with parameter $\theta_1$, and $u_{safe} = \tau(u_{nn}, \theta_2)=\tau(x, \theta_1, \theta_2)$ represents the output of the CBF-QP layer \eqref{eq:qp-u} with the parameter $\theta_2$. 
This allows real-time satisfaction of the safety requirements by dynamically modulating the control input. Additionally, the QP layer enables learning of the function $\alpha(\cdot)$, providing greater flexibility in enforcing safety constraints.
\item Training Objective: The control policy $\tau(x,\theta_1, \theta_2)$ is trained to minimize $\ell$ as in \eqref{eq:formulation}, which incorporates a CLF-based loss to stabilize the system while embedding CBFs as hard constraints in the QP layer to enforce safety. The differentiable framework enables gradient-based optimization while maintaining compliance with safety constraints.
\end{itemize}
The proposed framework is illustrated in Figure \ref{fig: structure}, which outlines the Neural ODE-based control architecture integrating a differentiable CBF-QP layer for safety enforcement.
\paragraph{Opt-ODENet: Gradients Computation} In the following, we show how to use the adjoint method to train the proposed framework in a memory efficient way.
\tikzset{every picture/.style={line width=0.75pt}} 
\begin{figure}[t]
\centering
\tikzset{every picture/.style={line width=0.75pt}} 

\begin{tikzpicture}[x=0.8pt,y=0.65pt,yscale=-1.1,xscale=1.1]

\draw  [color={rgb, 255:red, 155; green, 155; blue, 155 }  ,draw opacity=1 ][fill={rgb, 255:red, 155; green, 155; blue, 155 }  ,fill opacity=0 ] (82,52.8) .. controls (82,34.69) and (96.69,20) .. (114.8,20) -- (450.2,20) .. controls (468.31,20) and (483,34.69) .. (483,52.8) -- (483,151.2) .. controls (483,169.31) and (468.31,184) .. (450.2,184) -- (114.8,184) .. controls (96.69,184) and (82,169.31) .. (82,151.2) -- cycle ;
\draw  [color={rgb, 255:red, 245; green, 166; blue, 35 }  ,draw opacity=1 ][fill={rgb, 255:red, 155; green, 155; blue, 155 }  ,fill opacity=0.5 ] (104.73,167.98) -- (104.19,106.68) -- (133.27,115.11) -- (133.82,176.42) -- cycle ;
\draw  [color={rgb, 255:red, 245; green, 166; blue, 35 }  ,draw opacity=1 ][fill={rgb, 255:red, 155; green, 155; blue, 155 }  ,fill opacity=0.5 ] (120.73,167.48) -- (120.19,106.18) -- (149.6,114.61) -- (150.15,175.92) -- cycle ;
\draw  [color={rgb, 255:red, 245; green, 166; blue, 35 }  ,draw opacity=1 ][fill={rgb, 255:red, 155; green, 155; blue, 155 }  ,fill opacity=0.5 ] (153.73,163.98) -- (153.19,102.68) -- (182.27,111.11) -- (182.82,172.42) -- cycle ;
\draw  [color={rgb, 255:red, 245; green, 166; blue, 35 }  ,draw opacity=1 ][fill={rgb, 255:red, 245; green, 166; blue, 35 }  ,fill opacity=0.5 ] (120.19,106.18) -- (129.12,101.14) -- (158.36,109.63) -- (149.44,114.67) -- cycle ;
\draw  [color={rgb, 255:red, 245; green, 166; blue, 35 }  ,draw opacity=1 ][fill={rgb, 255:red, 245; green, 166; blue, 35 }  ,fill opacity=0.5 ] (158.7,170.42) -- (150.15,175.92) -- (149.81,115.13) -- (158.36,109.63) -- cycle ;
\draw   (71.83,149.08) -- (87.63,149.08) -- (87.63,148) -- (98.17,150.17) -- (87.63,152.33) -- (87.63,151.25) -- (71.83,151.25) -- cycle ;
\draw   (197.5,148.88) -- (216.1,148.88) -- (216.1,147.5) -- (228.5,150.25) -- (216.1,153) -- (216.1,151.63) -- (197.5,151.63) -- cycle ;
\draw  [color={rgb, 255:red, 74; green, 144; blue, 226 }  ,draw opacity=1 ][fill={rgb, 255:red, 74; green, 144; blue, 226 }  ,fill opacity=0.5 ] (239.33,137.83) .. controls (239.33,133.42) and (242.92,129.83) .. (247.33,129.83) -- (301.33,129.83) .. controls (305.75,129.83) and (309.33,133.42) .. (309.33,137.83) -- (309.33,161.83) .. controls (309.33,166.25) and (305.75,169.83) .. (301.33,169.83) -- (247.33,169.83) .. controls (242.92,169.83) and (239.33,166.25) .. (239.33,161.83) -- cycle ;
\draw   (361,137.67) .. controls (361,133.25) and (364.58,129.67) .. (369,129.67) -- (458.5,129.67) .. controls (462.92,129.67) and (466.5,133.25) .. (466.5,137.67) -- (466.5,161.67) .. controls (466.5,166.08) and (462.92,169.67) .. (458.5,169.67) -- (369,169.67) .. controls (364.58,169.67) and (361,166.08) .. (361,161.67) -- cycle ;
\draw   (269.7,101.49) -- (269.82,115.3) -- (270.92,115.29) -- (268.79,124.52) -- (266.51,115.33) -- (267.61,115.32) -- (267.49,101.51) -- cycle ;
\draw  [color={rgb, 255:red, 74; green, 144; blue, 226 }  ,draw opacity=1 ][fill={rgb, 255:red, 74; green, 144; blue, 226 }  ,fill opacity=0.5 ] (347,64.4) .. controls (347,59.76) and (350.76,56) .. (355.4,56) -- (469.1,56) .. controls (473.74,56) and (477.5,59.76) .. (477.5,64.4) -- (477.5,89.6) .. controls (477.5,94.24) and (473.74,98) .. (469.1,98) -- (355.4,98) .. controls (350.76,98) and (347,94.24) .. (347,89.6) -- cycle ;
\draw  [color={rgb, 255:red, 245; green, 166; blue, 35 }  ,draw opacity=1 ][fill={rgb, 255:red, 155; green, 155; blue, 155 }  ,fill opacity=0.5 ] (206.42,74.93) -- (206.19,48.98) -- (218.33,52.55) -- (218.55,78.5) -- cycle ;
\draw  [color={rgb, 255:red, 245; green, 166; blue, 35 }  ,draw opacity=1 ][fill={rgb, 255:red, 155; green, 155; blue, 155 }  ,fill opacity=0.5 ] (213.09,74.72) -- (212.87,48.77) -- (225.14,52.34) -- (225.37,78.29) -- cycle ;
\draw  [color={rgb, 255:red, 245; green, 166; blue, 35 }  ,draw opacity=1 ][fill={rgb, 255:red, 155; green, 155; blue, 155 }  ,fill opacity=0.5 ] (226.87,73.24) -- (226.64,47.29) -- (238.77,50.86) -- (239,76.81) -- cycle ;
\draw  [color={rgb, 255:red, 245; green, 166; blue, 35 }  ,draw opacity=1 ][fill={rgb, 255:red, 245; green, 166; blue, 35 }  ,fill opacity=0.5 ] (212.87,48.77) -- (216.59,46.64) -- (228.8,50.23) -- (225.07,52.37) -- cycle ;
\draw  [color={rgb, 255:red, 245; green, 166; blue, 35 }  ,draw opacity=1 ][fill={rgb, 255:red, 245; green, 166; blue, 35 }  ,fill opacity=0.5 ] (228.94,73.7) -- (225.37,76.03) -- (225.23,50.3) -- (228.8,47.98) -- cycle ;
\draw  [color={rgb, 255:red, 245; green, 166; blue, 35 }  ,draw opacity=1 ][fill={rgb, 255:red, 245; green, 166; blue, 35 }  ,fill opacity=0.5 ] (289.5,59.25) .. controls (289.5,57.46) and (290.96,56) .. (292.75,56) .. controls (294.54,56) and (296,57.46) .. (296,59.25) .. controls (296,61.04) and (294.54,62.5) .. (292.75,62.5) .. controls (290.96,62.5) and (289.5,61.04) .. (289.5,59.25) -- cycle ;
\draw   (175.5,54.88) -- (185.9,54.88) -- (185.9,54) -- (192.83,55.75) -- (185.9,57.5) -- (185.9,56.63) -- (175.5,56.63) -- cycle ;
\draw   (323,148.38) -- (341.6,148.38) -- (341.6,147) -- (354,149.75) -- (341.6,152.5) -- (341.6,151.13) -- (323,151.13) -- cycle ;
\draw   (470,148.38) -- (488.6,148.38) -- (488.6,147) -- (501,149.75) -- (488.6,152.5) -- (488.6,151.13) -- (470,151.13) -- cycle ;
\draw    (194,93.5) -- (331,93.5) ;

\draw (46.5,141.4) node [anchor=north west][inner sep=0.75pt]  [font=\footnotesize]  {$x(t_0)$};
\draw (35.17,106.33) node [anchor=north west][inner sep=0.75pt]  [font=\large] [align=center] {{\fontfamily{pcr}\selectfont {\tiny \textbf{Initial}}}\\[-5pt]{\fontfamily{pcr}\selectfont {\tiny \textbf{state}}}};
\draw (113.17,71.5) node [anchor=north west][inner sep=0.75pt]  [font=\large] [align=center] {{\fontfamily{pcr}\selectfont {\tiny \textbf{Controller}}}\\[-5pt]
{\fontfamily{pcr}\selectfont {\tiny \textbf{Network}}}
};
\draw (198.5,129.57) node [anchor=north west][inner sep=0.75pt]  [font=\footnotesize]  {$u_{nn}$};
\draw (243,138) node [anchor=north west][inner sep=0.75pt]   [align=center] {
{\fontfamily{pcr}\selectfont {\tiny \textbf{Differentiable}}}\\[-2pt]
{\fontfamily{pcr}\selectfont {\tiny \textbf{CBF-QP Layer}}}
};
\draw (324.17,129.73) node [anchor=north west][inner sep=0.75pt]  [font=\footnotesize]  {$u_{safe}$};
\draw (380,114.33) node [anchor=north west][inner sep=0.75pt] [font=\large]   [align=left] {{\fontfamily{pcr}\selectfont {\tiny \textbf{Plant Dynamics}}}};
\draw (364,143.57) node [anchor=north west][inner sep=0.75pt]  [font=\scriptsize]  {$\dot{x} =f( x) +g( x) u_{safe}$};
\draw (504.5,141.9) node [anchor=north west][inner sep=0.75pt]  [font=\footnotesize]  {$x(t_{f})$};
\draw (230.17,4) node [anchor=north west][inner sep=0.75pt]  [font=\large] [align=left] {{\fontfamily{pcr}\selectfont {\scriptsize \textbf{Opt-ODENet for Safety}}}};
\draw (387.67,60) node [anchor=north west][inner sep=0.75pt] [font=\large]  [align=left] {{\fontfamily{pcr}\selectfont {\tiny \textbf{ODE Solve}}}\\};
\draw (355,77.4) node [anchor=north west][inner sep=0.75pt]  [font=\scriptsize]  {$( x( t_{0}) ,\ f,\ g,\ [ t_{0} ,\ ...\ ,t_{f}])$};
\draw (288,105.4) node [anchor=north west][inner sep=0.75pt]  [font=\footnotesize]  {$\kappa $};
\draw (256.17,51.5) node [anchor=north west][inner sep=0.75pt]   [align=left] {{\fontfamily{pcr}\selectfont {\tiny \textbf{or}}}};
\draw (277.17,79) node [anchor=north west][inner sep=0.75pt]  [font=\footnotesize] [align=left] {{\fontfamily{pcr}\selectfont {\tiny \textbf{linear case}}}};
\draw (158,51.9) node [anchor=north west][inner sep=0.75pt]  [font=\tiny]  {$x(t)$};
\draw (202.17,79) node [anchor=north west][inner sep=0.75pt]  [font=\footnotesize] [align=left] {{\fontfamily{pcr}\selectfont {\tiny \textbf{general case}}}};
\draw (268,26) node [anchor=north west][inner sep=0.75pt]   [align=left] {{\fontfamily{pcr}\selectfont {\tiny \textbf{single parameter}}}};
\draw (179,26) node [anchor=north west][inner sep=0.75pt]   [align=left] {{\fontfamily{pcr}\selectfont {\tiny \textbf{function \ \ \ \ Network}}}};
\draw (213,25) node [anchor=north west][inner sep=0.75pt]  [font=\tiny]  {$\alpha(\cdot) $};

\end{tikzpicture}
\caption{Schematics of the Neural ODE-based controller with a differentiable CBF-QP layer enforcing safety constraints}
\vspace{-0.2cm}
\label{fig: structure}
\end{figure}
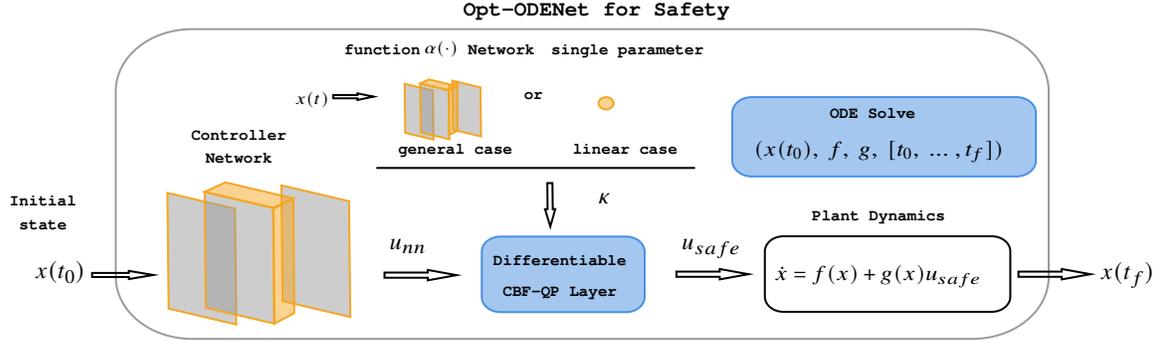
\begin{proposition}\label{proposition}
    Consider the Neural ODE-based framework \ref{fig: structure} and the loss function $\ell$. Let $\theta_1$ be the parameter of the controller neural network $\tau(\cdot)$, and $\theta_2$ the parameter of the differentiable CBF-QP layer. Define $\mu_1(t_0)$ and $\mu_2(t_0)$ by the gradient of loss $\ell$ to $\theta_1$ and $\theta_2$:
\begin{equation}   \mu_1(t_0):=\nabla_{\theta_1}\ell, \quad \mu_2(t_0):=\nabla_{\theta_2}\ell.
\end{equation}
Then, the gradients $\mu_1$ and $\mu_2$ can be updated by the following differential equations with boundary conditions where $p$ is a Lagrange multiplier:
\begin{align}
    &\dot x(t)=\mathcal{F}(x(t),\tau(x,\theta_1,\theta_2)),\quad x(t_0)=x_0,\nonumber\\
    &\dot p(t)=-p\frac{\partial \mathcal{F}}{\partial x}-\frac{\partial \mathcal{L}}{\partial x},\quad p(t_f)=\frac{\partial \Phi}{\partial x(t_f)},\nonumber\\
    &\dot \mu_1(t)=-p\frac{\partial \mathcal{F}}{\partial \theta_1}-\frac{\partial \mathcal{L}}{\partial \theta_1},\quad \mu_1(t_f)=\vmathbb{0}_{n_{\theta_1}},\label{eq:adjoint}\\
    &\dot \mu_2(t)=-p\frac{\partial \mathcal{F}}{\partial \theta_2}-\frac{\partial \mathcal{L}}{\partial \theta_2},\quad \mu_2(t_f)=\vmathbb{0}_{n_{\theta_2}},\nonumber
\end{align}
\end{proposition}
The differential equations \eqref{eq:adjoint} are derived by applying the adjoint method to an optimal control problem. In the following proof, we demonstrate the gradient computation for Neural ODEs with an embedded differentiable QP layer, leveraging the interconnected structure of the controller network. Due to space limitations, more details of the proof can be found in the \ref{app}.
\begin{proof}
Consider \eqref{eq: control-affine system}; we have that
\begin{equation}
    \frac{\partial \mathcal{F}}{\partial x}=\frac{\partial f}{\partial x}+\frac{\partial g}{\partial x}\tau(x,\theta_1,\theta_2)+g\frac{\partial \tau(x,\theta_1,\theta_2)}{\partial x}.
\end{equation}
Using chain rule, $\frac{\partial \tau(x,\theta_1,\theta_2)}{\partial x}$ is equal to
\begin{equation}
    \frac{\partial \tau(x,\theta_1,\theta_2)}{\partial x}=\frac{\partial \tau(x,\theta_1,\theta_2)}{\partial \pi(x, \theta_1)}\frac{\partial \pi(x, \theta_1)}{\partial x},
\end{equation}
where $\frac{\partial \pi(x, \theta_1)}{\partial x}$ can be calculated by Automatic Differentiation (AD).
We then calculate the first term by considering the first-order optimality condition of the differentiable CBF-QP layer \eqref{eq:qp-u}.
Consider the general case of QP problem where all components $A$, $b$, $G$, $h$, $Q$, and $q$ are present, and $Q$ is positive definite, $A$ and  $G$ have full rank . The Lagrangian of \eqref{eq:qp-u} is given by
\begin{equation}
    L(u,\nu, \lambda) =\frac{1}{2}u^{\top}Qu+q^{\top}u+\nu^{\top}(Au-b)+\lambda^{\top}(Gu-h)
\end{equation}
where $\nu$ are the dual variables on the equality constraints and $\lambda \geq 0$ are the dual variables on the inequality constraints. The KKT conditions for stationary, primal feasibility, and complementary slackness are
\begin{equation}
    \begin{aligned}
        Qu^* + q +A^{\top}\nu^* + G^{\top}\lambda^* &=0\\
        Au^*-b &=0\\
        D(\lambda^*)(Gu^*-h)&=0,
    \end{aligned}
    \label{KKT}
\end{equation}
where $D(\cdot)$ is a diagonal matrix with entries those of the vector argument, and $u^*$, $\nu^*$ and $\lambda^*$ are the optimal primal and dual variables. Taking the differentials of these conditions gives the equations
\begin{equation}
    \begin{bmatrix}
        Q & A^{\top}  & G^{\top}\\
        A & 0 & 0\\
        D(\lambda^*)G & 0 & D(Gu^*-h)
    \end{bmatrix}\begin{bmatrix}
        du\\
        d\nu\\
        d\lambda 
        \end{bmatrix}= -\begin{bmatrix}
            dQu^*+dq+dG^{\top}\lambda^*+dA^{\top}\nu^*\\
            dAu^*-db\\
            D(\lambda^*)dGu^*-D(\lambda^*)dh
            \end{bmatrix}
\end{equation}
First, according to \eqref{KKT}, we can have an implicit function $F$ \citep{dontchev2009implicit}:
\begin{equation}
    F(X, Y) = \begin{bmatrix}
        Qu^* + q + A^{\top} \nu^* +G^{\top}\lambda^*\\
        Au^* - b\\
        D(\lambda^*)(Gu^*-h)
    \end{bmatrix}=0
\end{equation}
where $X = [Q;q;A;b;G;h]$, $Y=[u;\nu;\lambda]$.
Hence, the Jacobian with respect to $Y$ is:
\begin{equation}\label{eq:19}
\begin{aligned}
    J_{F, Y} &= \left[\begin{array}{c|c|c}
        J_{F,u} & J_{F,\nu} &J_{F,\lambda}\end{array}\right]\\
        &=\left[\begin{array}{c|c|c}
         Q  & A^{\top} & G^{\top} \\
         A    & 0 & 0\\
         D(\lambda^*)G & 0 & D(Gu^*-h)
        \end{array}\right]
\end{aligned}
\end{equation}
note that all the vectors are column vectors. Next, we calculate the Jacobian with respect to $X$ (note that all the matrices and vectors are operated as column vectors):
\begin{equation}\label{eq:20}
\begin{aligned}
    J_{F, X} & = \left[\begin{array}{c|c|c|c|c|c}
      J_{F,Q}   & J_{F,q} & J_{F,A} & J_{F,b} & J_{F,G} & J_{F,h} 
    \end{array} \right]\\
    & = \left[\begin{array}{c|c|c|c|c|c}
      I\otimes u^{*\top}   & I & \nu^{*\top}\otimes I & 0 & \lambda^{*\top}\otimes I & 0\\
      0 & 0 & I\otimes u^{*\top} & -I &0 & 0\\
      0 & 0 & 0 & 0 & D(\lambda^*)I\otimes u^{*\top} & -D(\lambda^*)
    \end{array} \right]
\end{aligned}
\end{equation}
where $\otimes$ represents Kronecker product.
The Jacobian of $Y$ with respect to $X$ is then given by
\begin{equation}\label{eq:jacobianYX}
    J_{Y,X}=-[J_{F,Y}]^{-1}[J_{F,X}],
\end{equation}
where $J_{F,Y}$ is given in \eqref{eq:19}, and $J_{F,X}$ is given in \eqref{eq:20}. Note in our problem, $A$ is not present rather than being set to $0$. This does not affect the invertibility of $J_{F,Y}$. $\frac{\partial \tau(x,\theta_1,\theta_2)}{\partial \pi(x, \theta_1)}$ can be obtained via $\frac{\partial \tau(x,\theta_1,\theta_2)}{\partial q}$ from $J_{Y,X}$. We then consider $\frac{\partial \mathcal{F}}{\partial \theta_2}$; we have that
\begin{equation}
    \frac{\partial \mathcal{F}}{\partial \theta_2}=\frac{\partial \mathcal{F}}{\partial \tau(x,\theta_1,\theta_2)}\frac{\partial \tau(x,\theta_1,\theta_2)}{\partial \theta_2}=\underbrace{\frac{\partial \mathcal{F}}{\partial \tau(x,\theta_1,\theta_2)}}_{AD}\underbrace{\frac{\partial \tau(x,\theta_1,\theta_2)}{\partial h(x, \theta_2)}}_{\eqref{eq:jacobianYX}}\underbrace{\frac{\partial h(x, \theta_2)}{\partial \theta_2}}_{AD}
\end{equation}
Similarly, for $\frac{\partial \mathcal{F}}{\partial \theta_1}$, it holds that
\begin{equation}
    \frac{\partial \mathcal{F}}{\partial \theta_1}=\frac{\partial \mathcal{F}}{\partial \tau(x,\theta_1,\theta_2)}\frac{\partial \tau(x,\theta_1,\theta_2)}{\partial \theta_1}=\underbrace{\frac{\partial \mathcal{F}}{\partial \tau(x,\theta_1,\theta_2)}}_{AD}\underbrace{\frac{\partial \tau(x_1,\theta_1,\theta_2)}{\partial \pi(x, \theta_1)}}_{\eqref{eq:jacobianYX}}\underbrace{\frac{\partial \pi(x, \theta_1)}{\partial \theta_1}}_{AD}.
\end{equation}
The remaining terms, $\frac{\partial \mathcal{L}}{\partial x}$, $\frac{\partial \mathcal{L}}{\partial \theta_1}$ and $\frac{\partial \mathcal{L}}{\partial \theta_2}$ can also be efficiently calculated via AD and \eqref{eq:jacobianYX} similarly as $\mathcal{L}$ is already an explicit function of $x$.
\end{proof}
\SetKwComment{Comment}{/* }{ */}
\LinesNumbered
\begin{algorithm2e}[t]
\caption{Training Algorithm}\label{alg: node-cbfqp-clf}
\KwIn{Sampled initial state $x(t_0)$, number of iterations $n > 0$, time steps $[t_0, t_1, \cdots, t_f]$, dynamics $f$ and $g$, safety requirements}
\KwOut{A safe and stable controller $u_{safe}$ and class-$\mathcal{K}$ function}
Construct CBF and CLF;\\
\While{$i \leq n$}{
$u_{nom} \gets NN_{controller}(x, \theta)$\;
$u_{safe} \gets QP(CBF, u_{nom})$\;
$x \gets ODESolver(x(t_0), f, g, u_{safe}, [t_0, \cdots, t_f])$
\;
Compute loss $\ell$ by CLF over the trajectories;\\
$\theta_1 \gets Optimizer(\nabla_{\theta_1} \ell, \theta_1)$ \Comment*[r]{Update controller NN parameters}
$\theta_2 \gets Optimizer(\nabla_{\theta_2} \ell, \theta_2)$ \Comment*[r]{Update function $\alpha(\cdot)$ parameters}
}
\end{algorithm2e}
\vspace{-0.5cm}
\paragraph{Control Objective: CLF Loss}
Inspired by LyaNet \cite{LyaNet}, for a given target and terminal loss $\Phi$,  the potential function $V$ can be designed as 
\begin{equation}
    V_{x}(\cdot) \coloneqq \Phi\left(x\left(\cdot \right)\right)
\end{equation}
Then a point-wise Lyapunov loss can be designed as
\begin{equation}
    \mathcal{V} \coloneqq \max \left\{0, \frac{\partial V_{x}}{\partial x}\mathcal{F} \left(x, u\right)+ \gamma\ V_{x}\left(x\right)\right\}
    \label{pwL}
\end{equation}
Equation \eqref{pwL} signifies the local violation of the invariance condition specified in \eqref{eq:clf}. When $\mathcal{V}=0$ holds for all data in the time interval, the inference dynamics exhibit exponential convergence towards a prediction that minimizes the loss. The Lyapunov loss for the dynamic system \eqref{eq: control-affine system} is 
\begin{equation}
    \ell \coloneqq \mathbb{E} \left[\int_{t_0}^{t_f} \mathcal{V}dt\right]
\end{equation}
\begin{remark}[\cite{LyaNet}]
Consider the Lyapunov loss above. If there exists a parameter $\theta^*$ of the dynamic system that satisfies $\ell(\theta^*)=0$, then:
\begin{itemize}
    \item The potential function $V_{x}$ is an exponentially stabilizing Lyapunov function with $\theta ^*$;
    \item For $t\in \left[t_0, t_f\right]$, the dynamics satisfy the following convergence expression with respect to the loss $\Phi$:
    \begin{equation}
            \Phi \left(x(t)\right) \leq \Phi \left(x(t_0)\right)e^{-\kappa t}
    \end{equation}
\end{itemize}
\end{remark}
The Lyapunov method provides a guaranteed convergence rate for a broader range of problems and affords us the opportunity to manually select the rate of convergence, as there might be instances where too fast dynamics are not preferred.\\
Algorithm \ref{alg: node-cbfqp-clf} demonstrates how the steps above are integrated in our proposed method. 
\section{Case Study}
\begin{figure}[t]
    \centering
    \begin{minipage}[t]{0.48\linewidth}
        \centering
        \includegraphics[height=4.5cm]{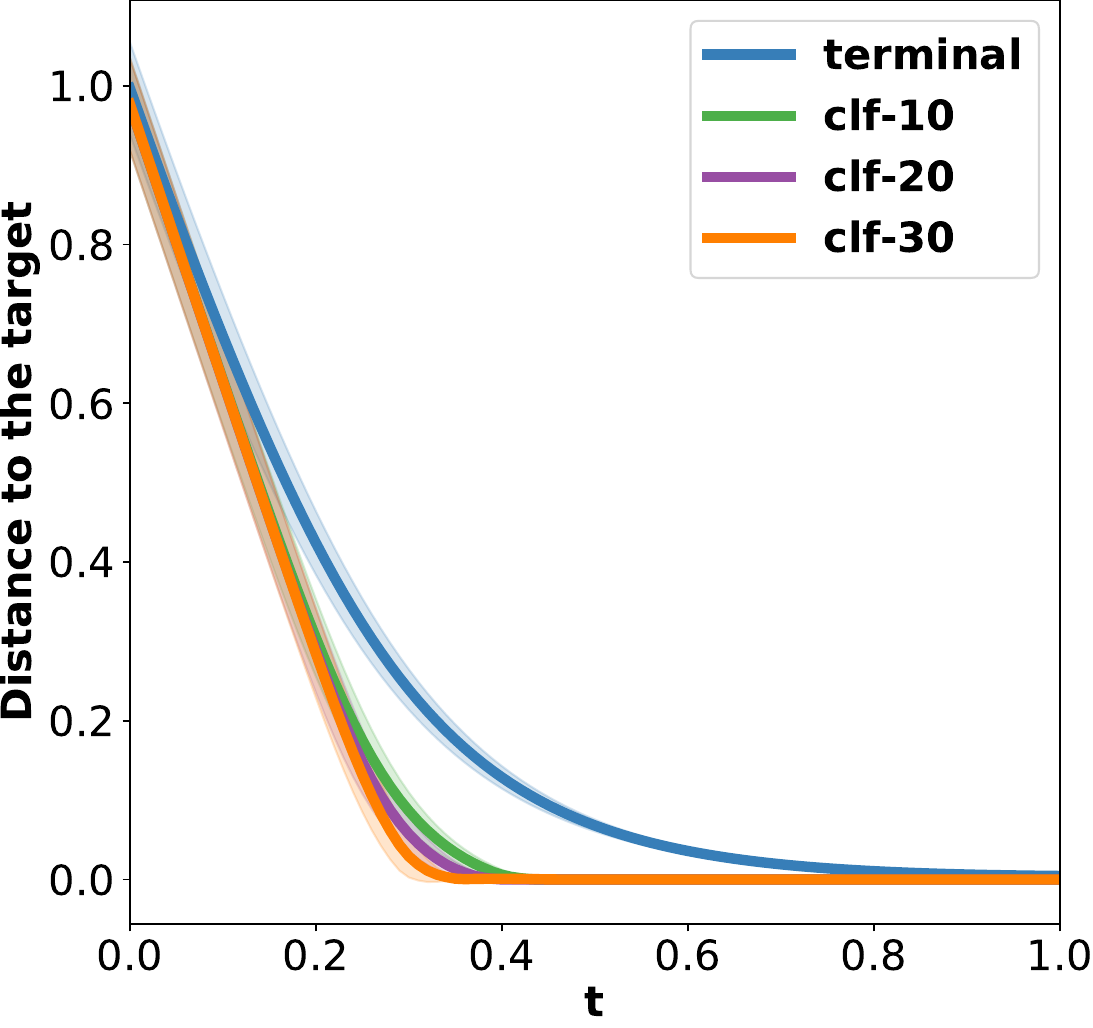}
        \vspace{-0.3cm}
        \caption{Distance over time with varying $\ell$}
        \label{fig:lossviatime-clf}
    \end{minipage}
    \begin{minipage}[t]{0.48\linewidth}
        \centering
        \includegraphics[height=4.5cm]{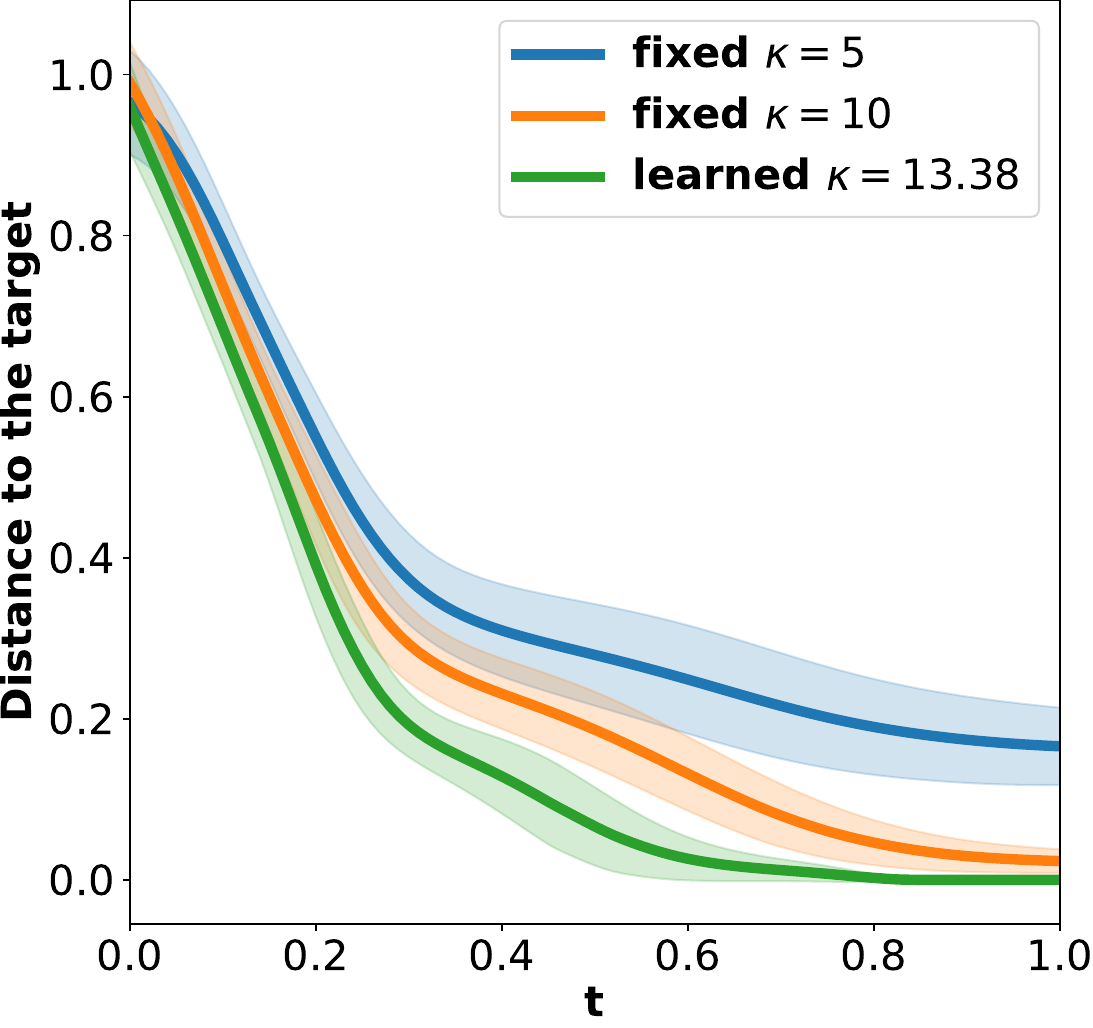}
        \vspace{-0.3cm}
        \caption{Distance over time with varying $\kappa$}
        \label{fig:lossviatime-kappa}
    \end{minipage}
\end{figure}
\begin{figure}[t]
    \centering
    \subfigure[without CBF-QP]{
    \begin{minipage}[t]{0.33\linewidth}
        \centering
        \includegraphics[height=2.8cm]{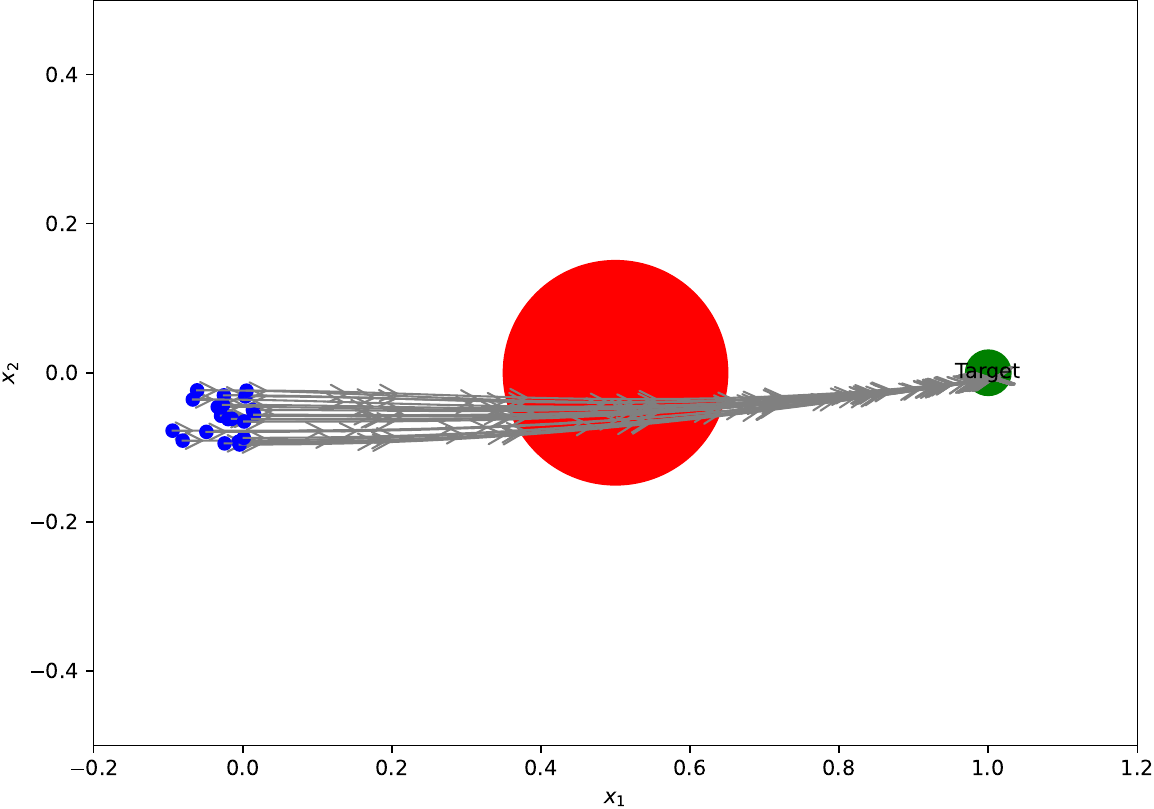}
    \end{minipage}\label{kappa1}}
    \hspace{-5mm}
    \subfigure[CBF-QP only in Inference]{
    \begin{minipage}[t]{0.33\linewidth}
        \centering
        \includegraphics[height=2.8cm]{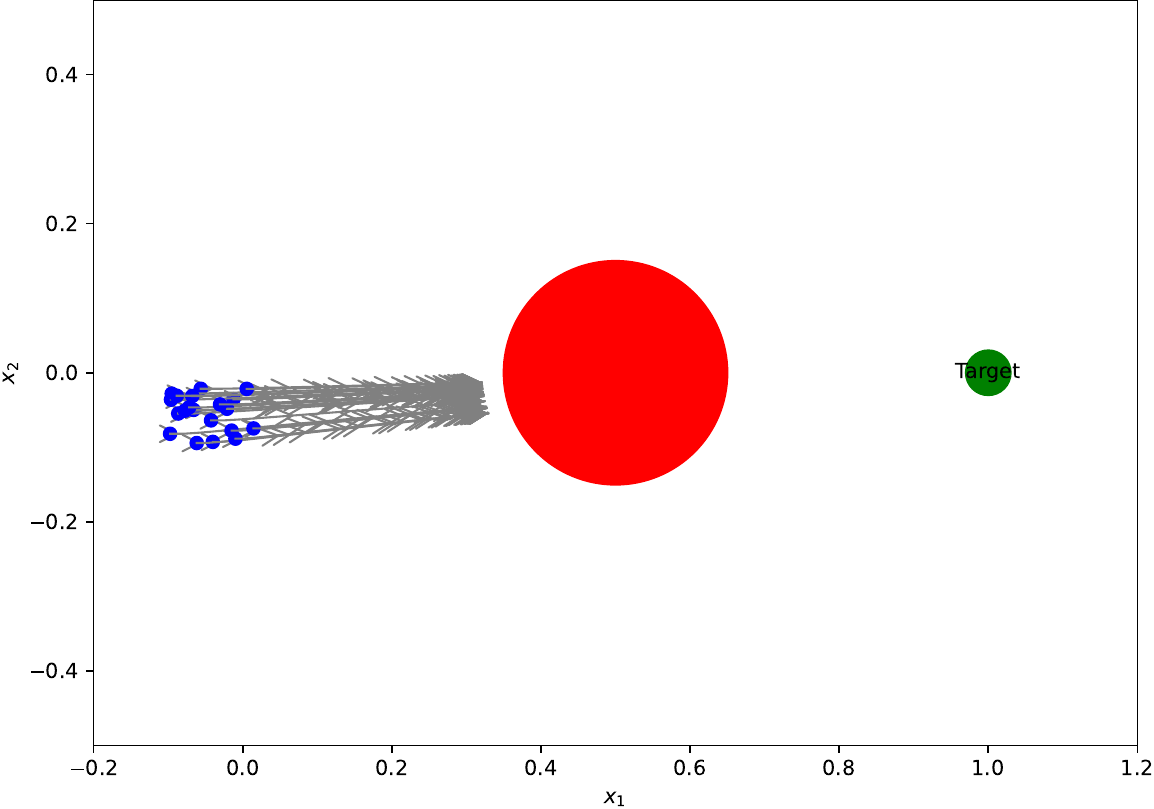}
    \end{minipage}\label{kappa10}}
    \hspace{-5mm}\\
    \subfigure[CBF-QP: fixed $\kappa=5$]{
    \begin{minipage}[t]{0.33\linewidth}
        \centering
        \includegraphics[height=2.8cm]{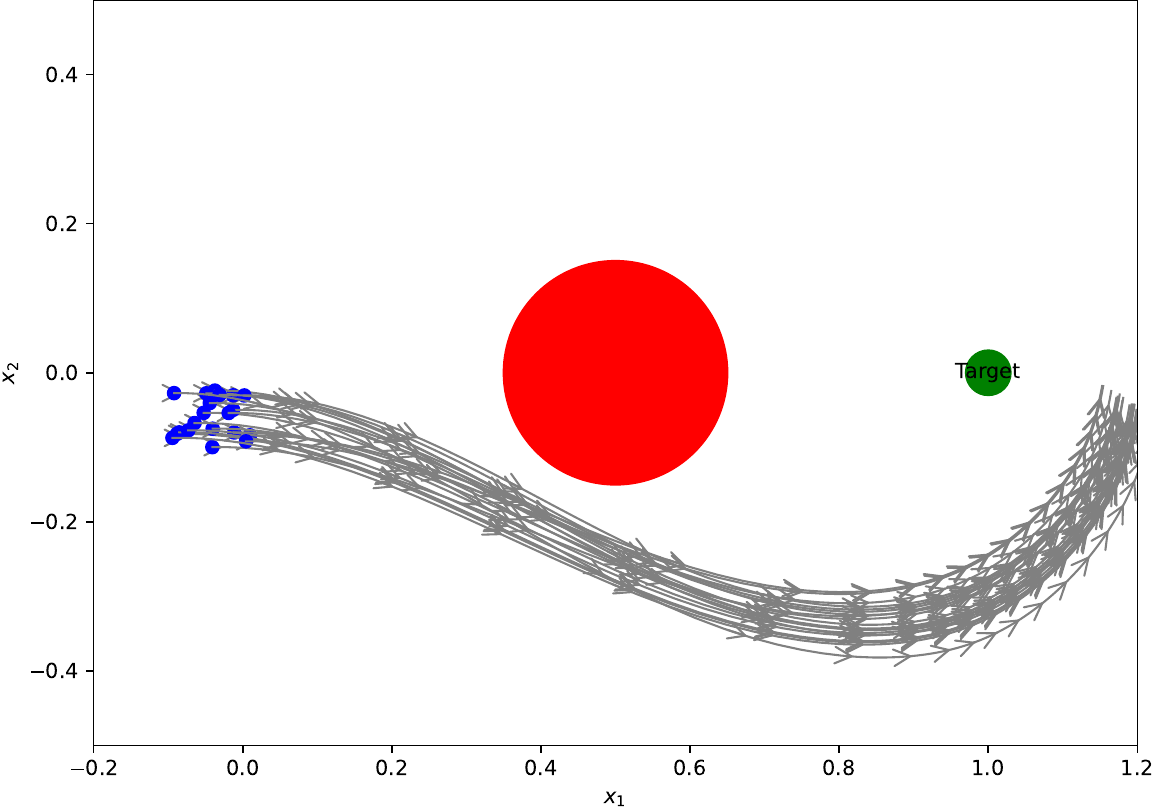}
    \end{minipage}\label{kappa20}}
    \hspace{-5mm}
    \subfigure[CBF-QP: fixed $\kappa=10$]{
    \begin{minipage}[t]{0.33\linewidth}
        \centering
        \includegraphics[height=2.8cm]{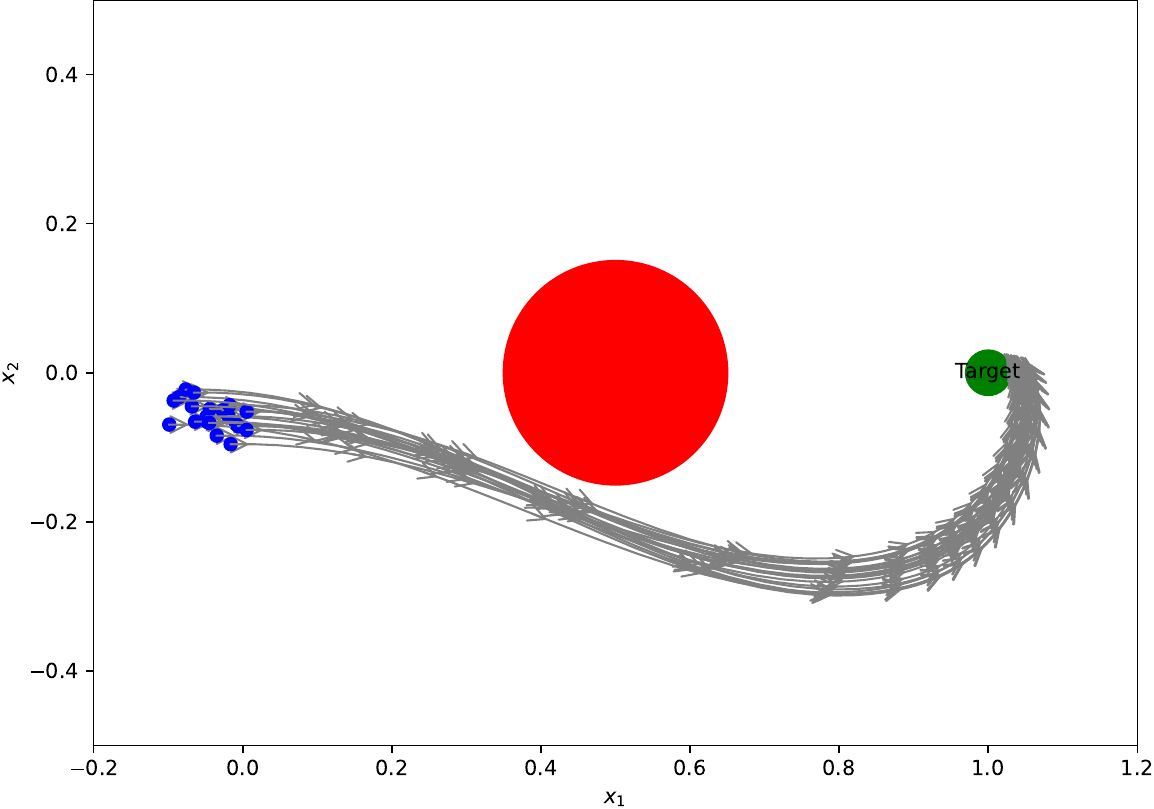}\
    \end{minipage}\label{kappa20-0.05}}
    \hspace{-5mm}
    \subfigure[CBF-QP: learned $\kappa=13.38$]{
    \begin{minipage}[t]{0.33\linewidth}
        \centering
        \includegraphics[height=2.8cm]{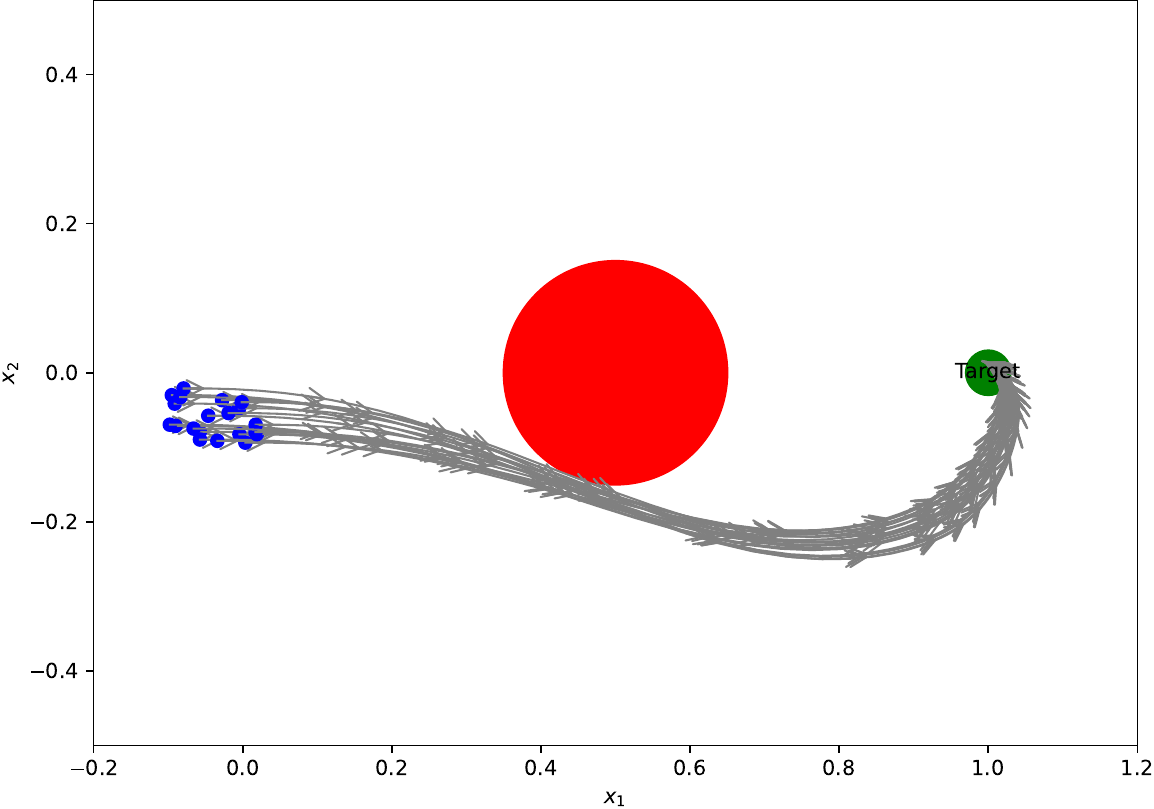}
    \end{minipage}\label{mnode}}
    \caption{Tested Trajectories under Different CBF-QP Settings}
    \label{fig:traj}
    \vspace{-0.5cm}
\end{figure}
\begin{figure}
    \centering
    \includegraphics[width=1.0\linewidth]{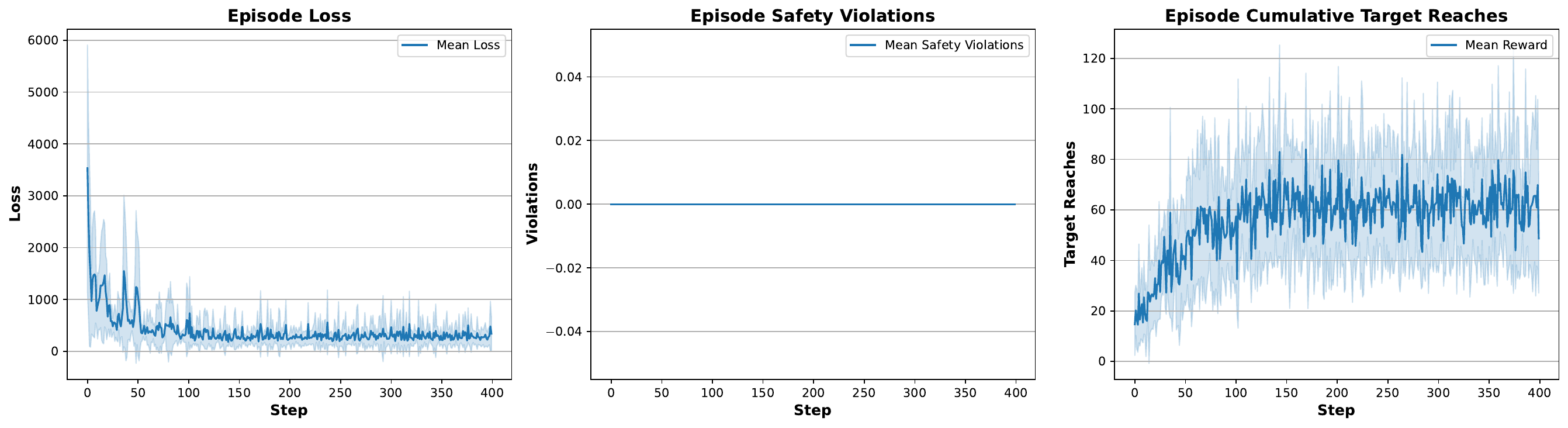}
    \caption{Training results on Simulated Cars with HOCBF}
    \label{fig:hocbf}
    \vspace{-0.5cm}
\end{figure}

\begin{table}[t]
\begin{tabular}{c|c|c|c|c|c}
\hline
          & fixed $\kappa$=5 & fixed $\kappa=10$ & learned $\kappa=13.38$ & No CBF-QP & \multicolumn{1}{l}{CBF-QP only in Inference} \\ \hline
Mean Error     & 0.3623           & 0.2544            & 0.2036                 & 0.1587    & 0.7088                                       \\ \hline
Collision & No               & No                & No                     & Yes       & No                                           \\ \hline
\end{tabular}
\caption{Comparison of Performance under Different Settings}
\label{table:compare}
\end{table}
In this experiment setup, a unicycle is tasked with reaching the designated location, i.e., the destination, while avoiding collisions with an obstacle. The dynamics of the system are:
\begin{equation}
        \dot{x}(t)=
        \left[ \begin{array}{cc}
        \cos \vartheta(t)  & 0 \\
        \sin \vartheta(t) & 0 \\
        0 & 1.0 
        \end{array}
        \right ] u(t).
        \label{eq:unicycle-dynamics}
\end{equation}
where $x = [x_1, x_2, \vartheta]^{\top}$ is the state. The control input is given by $u(t) = \tau(x(t); \theta_1, \theta_2)$ trained using our framework. A point at a distance $l_p$ ahead of the unicycle is used to establish safety constraints for collision-free navigation, and the function $p:\mathbb{R}^3 \rightarrow \mathbb{R}^2$ is defined as:
\begin{equation}
        p(x(t))=
        \left[ \begin{array}{c}
                x_1(t) \\
                x_2(t) \\
        \end{array} 
        \right ]+l_p
        \left[ \begin{array}{c}
                \cos \vartheta(t) \\
                \sin \vartheta(t) \\
        \end{array}
        \right].
        \label{eq:unicycle-modify}
\end{equation}
In this case, the CBF is defined as $h(x(t)) = \frac{1}{2} \big((p(x(t)) - p_{\text{obs}})^2 - \delta_1^2 \big)$ , where  $p_{\text{obs}}$ represents the position of the obstacle, and $\delta_1$ denotes the minimum safe distance between the unicycle and the obstacle. For $\alpha(\cdot)$, we adopt a linear form where only the linear coefficient $\kappa$ is learned, as illustrated in Figure \ref{fig: structure}. The CLF is defined as  $V(x(t)) = \frac{1}{2} \big((p(x(t)) - p_{\text{tar}})^2 - \delta_2^2 \big)$, where  $p_{\text{tar}}$  represents the position of the target, and  $\delta_2$  specifies the radius of the target region.\\
First, we compared the system’s performance when only a terminal loss was used versus when a CLF-based loss was incorporated. The trajectories were evaluated by tracking the evolution of the system’s distance to the target over time, as shown in Figure \ref{fig:lossviatime-clf}. The curve with the CLF (with varying $\gamma$) converges significantly faster to the target compared to the one with only terminal loss as in \cite{bottcher2022ai}, highlighting the advantage of CLF in accelerating convergence.\\
Next, we analyzed the role of the CBF-QP layer in ensuring safety and achieving the task. Several configurations were tested, including no QP layer as in \cite{ip2024lyapunov}, applying the QP layer only during inference (i.e., without being involved during controller training), and using fixed or learned $\alpha(\cdot)$ parameters as in \cite{pereira2021safe} (in our case, the parameter is the linear coefficient $\kappa$).\\
The results, shown in Figure \ref{fig:traj}, reveal significant differences in performance. Without the QP layer, the system fails to avoid obstacles and results in collisions. When the QP layer is applied only during inference, safety is maintained (in contrast, the method in \cite{sandoval2023neural} using soft constraints lacks theoretical safety guarantees), but the system loses its ability to effectively approach the target. For fixed $\kappa = 5$ and $\kappa = 10$, the method either leads to overly conservative behaviour, preventing task completion, or fails to strike a balance between safety and performance. In contrast, learning $\kappa$ during training enables the system to adaptively balance safety and task objectives.\\
Figure \ref{fig:lossviatime-kappa} shows  the evolution of the distance to the target over time with different coefficient $\kappa$. The presence of obstacles naturally slow convergence due to safety constraints compared to Figure \ref{fig:lossviatime-clf}. However, learning $\kappa$ achieves a better trade-off, ensuring collision-free navigation while maintaining effective convergence. Table \ref{table:compare} reports the mean error and the results of collision, showing that without the CBF-QP layer, collisions occur, inference-only QP sacrifices performance, and fixed $\kappa$ is either too conservative or fails. The learned $\kappa$ achieves minimal error with no collisions.\\
We also extend our framework to the High Order CBF (HOCBF) \citep{xiao2019control} case in the Simulated Cars environment, with Figure \ref{fig:hocbf} presenting the training results. The results demonstrate that the system successfully converges to task objectives while satisfying safety constraints. Additional comparisons, analysis and experiments details are shown in the \ref{app}.
\vspace{-0.6cm}
\section{Discussions and Future work}
Future work will focus on several key areas. First, designing more general and effective class-$\mathcal{K}$ functions for the CBF can improve the framework’s flexibility and performance. Second, optimizing the computational efficiency of the QP layer is crucial for enabling real-time applications. Third, extending the framework to tackle more complex tasks, such as handling moving obstacles, will further validate its robustness. Finally, exploring learning-based methods to directly design the CBF could lead to a fully adaptive and scalable control framework.
\acks{The authors would like to express their sincere gratitude for financial support from the Engineering and Physical Sciences Research Council (EPSRC): Keyan Miao was supported through the grant EP/T517811/1, and Antonis Papachristodoulou through the EEBio Programme Grant EP/Y014073/1.}
\bibliography{reference}
\section{Appendix}
\label{app}
\subsection{Gradients Computation of Opt-ODENet}
\subsubsection{Adjoint Method for Back-propagation in Neural ODEs}
A Neural ODE is a system of the form
\begin{equation}
\dot {x} \left(t\right) = \mathcal{F}\left(t, x\left(t\right), \theta\right) \quad t \in \mathcal{S}
\label{node-app}
\end{equation}
where $\mathcal{S}:= \left[t_0, t_f\right]$ ($t_0, t_f\in \mathbb{R}^+$) is the depth domain and $\mathcal{F}$ is a neural network called ODENet which is chosen as part of the machine learning model with parameter $\theta$.\\
Loss:
\begin{equation}
\ell \coloneqq \Phi\left(x \left(t_f\right)\right) + \int_{t_0}^{t_f}\mathcal{L}\left(x\left(t\right),\theta,t\right)dt
\label{int}
\end{equation}
Consider Problem \eqref{node-app}-\eqref{int}. The gradient of loss $\ell$ with respect to parameter $\theta$ is
\begin{equation}
\nabla_\theta \ell = \mu \left(t_0\right)
\label{gradl}
\end{equation}
where $x\left(t\right)$, $p\left(t\right)$ and $\mu\left(t\right)$ satisfy the boundary value problem:
\begin{equation}
\begin{aligned}
& \dot {x}\left(t\right) = \mathcal{F}, x\left(t_0\right) = x_0\\
& \dot p\left(t\right) = -p\left(t\right) \frac{\partial \mathcal{F}}{\partial x} - \frac{\partial \mathcal{L}}{\partial z}, \; p\left(t_f\right) = \frac{\partial \Phi}{\partial x\left(t_f\right)}\\
& \dot {\mu} \left(t\right) = -p\left(t\right)\frac{\partial \mathcal{F}}{\partial \theta} - \frac{\partial \mathcal{L}}{\partial \theta}, \; \mu \left(t_f\right) = \vmathbb{0}_{n_{\theta}}
\end{aligned}
\end{equation}
\begin{proof}
Define the augmented objective function with a Lagrange multiplier $p$ as
\begin{equation}
L = \ell - \int_{t_0}^{t_f} p(t)\left[\dot x(t) - \mathcal{F}\left(t, x(t), \theta\right)\right] dt
\label{augL}
\end{equation}
and $\dot x - \mathcal{F} =0$ always holds by construction, so $p(t)$ can be freely assigned while $\frac{d L}{d \theta} = \frac{d \ell}{d \theta}$. As to the integration part on right hand side of \eqref{augL}, we have
$$
\begin{aligned}
\int_{t_0}^{t_f}p(t)\left(\dot x - \mathcal{F}\right)dt & = p(t)x(t)\big{|}_{t_0}^{t_f}\\ 
& - \int_{t_0}^{t_f}\dot p(t)x(t)dt - \int_{t_0}^{t_f}p(t)\mathcal{F}dt\\
& = p\left(t_f\right)x\left(t_f\right)\\ 
& - p\left(t_0\right)x\left(t_0\right) - \int_{t_0}^{t_f}\left(\dot p(t)x(t) + p(t)\mathcal{F}\right)dt
\end{aligned}
$$
Hence,
$$
\begin{aligned}
L & = \Phi\left(x\left(t_f\right)\right)- p\left(t_f\right)x\left(t_f\right) + p\left(t_0\right)x\left(t_0\right)\\
& + \int_{t_0}^{t_f}\left(\dot p(t)x(t) + p(t)\mathcal{F} + \mathcal{L} \right)dt
\end{aligned}
$$
Then the gradient of $\ell$ with respect to $\theta$ can be computed as
\begin{equation}
\begin{aligned}
\frac{d \ell}{d \theta} & = \frac{d L}{d \theta} = \left(\frac{\partial \Phi}{\partial x\left(t_f\right)}-p(t_f)\right)\frac{d x(t_f)}{d \theta} \\
& + \int_{t_0}^{t_f}\left(\dot p(t)\frac{d x(t)}{d \theta} + p(t)\left(\frac{\partial \mathcal{F}}{\partial \theta} + \frac{\partial \mathcal{F}}{\partial x}\frac{d x}{d \theta}\right) + \frac{\partial \mathcal{L}}{\partial \theta} + \frac{\partial \mathcal{L}}{\partial x}\frac{d x}{d \theta}\right)dt
\end{aligned}
\label{proof-gradl}
\end{equation}
Now if we set this Lagrange multiplier as the adjoint state for the Hamiltonian function
\begin{equation}
H\left(x,p,\theta\right) = p\mathcal{F} + \mathcal{L}
\end{equation}
and according to PMP, the optimality conditions requires 
\begin{equation}
\dot x = \frac{\partial H}{\partial p} = \mathcal{F},\ \dot p = -\frac{\partial H}{\partial x} = -p\frac{\partial \mathcal{F}}{\partial zx} - \frac{\partial \mathcal{L}}{\partial x}
\label{PMP-condition}
\end{equation}
with initial conditions $x(t_0) = x_0$ and $p(t_f)=\frac{\partial \Phi}{\partial x(t_f)} = \frac{\partial \Phi}{\partial x\left(t_f\right)}$.
Substituting \eqref{PMP-condition} into \eqref{proof-gradl}, it can be obtained that
\begin{equation}
\begin{aligned}
\frac{d \ell}{d \theta} & = \int_{t_0}^{t_f}\left(p(t)\frac{\partial \mathcal{F}}{\partial \theta} + \frac{\partial L}{\partial \theta}\right)dt\\
& = \int_{t_f}^{t_0}\left(- p(t)\frac{\partial \mathcal{F}}{\partial \theta} - \frac{\partial \mathcal{L}}{\partial \theta}\right)dt = \int_{t_f}^{t_0} -\frac{\partial H}{\partial \theta}dt
\end{aligned}
\end{equation}
proving the result.
\end{proof}
\subsubsection{Gradients Through QP Layer}
For a differentiable CBF-QP layer \eqref{eq:qp-u}, the gradients of loss with respect to different variables through QP layer can be computed as follows.
\begin{theorem}[Implicit Function Theorem]
    Given three open setx $X \subseteq \mathbb R^n$, $Y \subseteq \mathbb R^m$ and $Z \subseteq \mathbb R^m$, if function $F: X \times Y \mapsto Z$ is continuously differentiable, and $(\hat x, \hat y)\in \mathbb R^n \times \mathbb R^m$ is a point for which
    \begin{equation}
        F(\hat x, \hat y) = \hat z,
    \end{equation}
    and the Jacobian of $F$ with respect to $y\subseteq Y$
    \begin{equation}
        J_{F,y}\bigg|_{i,j} = \left[\frac{\partial F_i}{\partial y_j}\right]
    \end{equation}
    is invertible at $(\hat x, \hat y)$, then there exists an open set $W\in \mathbb R^n$ with $x\in W$ and a unique continuously differentiable function $\phi: W \mapsto Y$ such that $y=\phi(x)$ and
    \begin{equation}
        F(x,y) = \hat z
    \end{equation}
    holds for $x\in W$.\\
    In addition, it can be shown that the partial derivatives of $\phi$ and $W$ are given by
    \begin{equation}
        J_{y,x} = -\left[J_{F,y}\right]^{-1}\left[J_{F,x}\right].
    \end{equation}
    \label{implicit-thm}
\end{theorem}
The Lagrangian of \eqref{eq:qp-u} is given by
\begin{equation}
    L(u,\nu, \lambda) =\frac{1}{2}u^{\top}Qu+q^{\top}u+\nu^{\top}(Au-b)+\lambda^{\top}(Gu-h)
\end{equation}
where $\nu$ are the dual variables on the equality constraints and $\lambda \geq 0$ are the dual variables on the inequality constraints. The KKT conditions for stationary, primal feasibility, and complementary slacknesss are
\begin{equation}
    \begin{aligned}
        Qu^* + q +A^{\top}\nu^* + G^{\top}\lambda^* &=0\\
        Au^*-b &=0\\
        D(\lambda^*)(Gu^*-h)&=0,
    \end{aligned}
    \label{KKT-app}
\end{equation}
where $D(\cdot)$ is a diagonal matrix with entries those of the vector argument, and $u^*$, $\nu^*$ and $\lambda^*$ are the optimal primal and dual variables. Taking the differentials of these conditions gives the equations
\begin{equation}
    \begin{aligned}
        dQu^* + Qdu + dq +dA^{\top}\nu^* + A^{\top} d\nu + dG^{\top}\lambda^* + G^{\top}d\lambda &=0\\
        dAu^* + Adu - db &=0\\
        D(Gu^*-h)d\lambda + D(\lambda^*)(dGu^* +Gdu -dh)&=0,
    \end{aligned}
\end{equation}
or written more compactly in matrix form
\begin{equation}
    \begin{bmatrix}
        Q & A^{\top}  & G^{\top}\\
        A & 0 & 0\\
        D(\lambda^*)G & 0 & D(Gu^*-h)
    \end{bmatrix}\begin{bmatrix}
        du\\
        d\nu\\
        d\lambda 
        \end{bmatrix}= -\begin{bmatrix}
            dQu^*+dq+dG^{\top}\lambda^*+dA^{\top}\nu^*\\
            dAu^*-db\\
            D(\lambda^*)dGu^*-D(\lambda^*)dh
            \end{bmatrix}
\end{equation}
First, according to \eqref{KKT-app}, we can have an implicit function $F$ \citep{dontchev2009implicit}:
\begin{equation}
    F(X, Y) = \begin{bmatrix}
        Qu^* + q + A^{\top} \nu^* +G^{\top}\lambda^*\\
        Au^* - b\\
        D(\lambda^*)(Gu^*-h)
    \end{bmatrix}=0
\end{equation}
where $X = [Q;q;A;b;G;h]$, $Y=[u;\nu;\lambda]$.
Hence, the Jacobian with respect to $Y$ is:
\begin{equation}\label{eq:47}
\begin{aligned}
    J_{F, Y} &= \left[\begin{array}{c|c|c}
        J_{F,u} & J_{F,\nu} &J_{F,\lambda}\end{array}\right]\\
        &=\left[\begin{array}{c|c|c}
         Q  & A^{\top} & G^{\top} \\
         A    & 0 & 0\\
         D(\lambda^*)G & 0 & D(Gu^*-h)
        \end{array}\right]
\end{aligned}
\end{equation}
note that all the vectors are column vectors. Next, we calculate the Jacobian with respect to $X$ (note that all the matrices and vectors are operated as column vectors):
\begin{equation}\label{eq:48}
\begin{aligned}
    J_{F, X} & = \left[\begin{array}{c|c|c|c|c|c}
      J_{F,Q}   & J_{F,q} & J_{F,A} & J_{F,b} & J_{F,G} & J_{F,h} 
    \end{array} \right]\\
    & = \left[\begin{array}{c|c|c|c|c|c}
      I\otimes u^{*T}   & I & \nu^{*T}\otimes I & 0 & \lambda^{*T}\otimes I & 0\\
      0 & 0 & I\otimes u^{*T} & -I &0 & 0\\
      0 & 0 & 0 & 0 & D(\lambda^*)I\otimes u^{*T} & -D(\lambda^*)
    \end{array} \right]
\end{aligned}
\end{equation}
where $\otimes$ represents Kronecker product.
According to the Implicit Function Theorem \ref{implicit-thm}, we obtain
\begin{equation}
\begin{aligned}
    \left(\frac{\partial l}{\partial X}\right)^{\top} &= \left(\frac{\partial l}{\partial Y}\right)^{\top}J_{Y, X}\\
    & = -\left(\frac{\partial l}{\partial Y}\right)^{\top} \left[J_{F,Y}\right]^{-1}\left[J_{F,X}\right]
\end{aligned}
\end{equation}
Since primal and dual variables $\lambda^*$ and $\nu^*$ will not be passed into following layers of networks, we have $\left(\frac{\partial l}{\partial Y}\right)^{\top}=\left[\left(\frac{\partial l}{\partial u^*}\right)^{\top}, 0, 0\right]$ where $0$ on the right hand side are zero vectors with the same size as $\nu^{*T}$ and $\lambda^{*T}$. We define$\begin{bmatrix}
        d_u\\
        d_{\nu}\\
        d_{\lambda}
    \end{bmatrix} = \left[J^{\top}_{F, Y}\right]^{-1}\frac{\partial l}{\partial Y}$. According to the Implicit Function Theorem, we obtain
$\frac{\partial l}{\partial X} = \left[J_{F,X}\right]^{\top}\begin{bmatrix}
        d_u\\
        d_{\nu}\\
        d_{\lambda}
    \end{bmatrix}$, then
\begin{equation}
\begin{aligned}
    \frac{\partial l}{\partial X} &= \left[\begin{array}{c|c|c|c|c|c}
      I\otimes u^{*T}   & I & \nu^{*T}\otimes I & 0 & \lambda^{*T}\otimes I & 0\\
      0 & 0 & I\otimes u^{*T} & -I &0 & 0\\
      0 & 0 & 0 & 0 & D(\lambda^*)I\otimes u^{*T} & -D(\lambda^*)
    \end{array} \right]^{\top} \begin{bmatrix}
        d_u\\
        d_{\nu}\\
        d_{\lambda}
    \end{bmatrix}\\
    & = \begin{bmatrix}
        I\otimes u^{*} & 0 & 0\\
        I & 0 & 0\\
        \nu^{*}\otimes I & I\otimes u^{*} & 0\\
        0 & -I & 0\\
        \lambda^{*}\otimes I & 0 & D(\lambda^*)I\otimes u^{*}\\
        0 & 0 & D(\lambda^*)
    \end{bmatrix}\begin{bmatrix}
        d_u\\
        d_{\nu}\\
        d_{\lambda}
    \end{bmatrix}
\end{aligned}
\end{equation}
Now, we can have
\begin{equation}\label{eq:qp-grad}
    \begin{aligned}
        \frac{\partial l}{\partial Q} & = \frac{1}{2}\left(d_u u^{*T} + u^*d_u^{\top}\right), \quad \frac{\partial l}{\partial q} = d_u\\
        \frac{\partial l}{\partial A} & = \nu^* d_u^{\top} + d_\nu u^{*T}, \quad \frac{\partial l}{\partial b} = -d_\nu\\
        \frac{\partial l}{\partial G} & = D(\lambda^*)d_{\lambda}z^{*T}+\lambda^*d_u^{\top}, \quad \frac{\partial l}{\partial h} = -D(\lambda^*)d_\lambda
    \end{aligned}
\end{equation}
Or more directly, for example, we can obtain
\begin{equation}
\begin{aligned}
    \left(\frac{\partial l}{\partial q}\right)^{\top} &= \left(\frac{\partial l}{\partial Y}\right)^{\top}[J_{F,Y}]^{-1}[J_{F,q}]\\
    &=\left(\frac{\partial l}{\partial u^*}\right)^{\top}[J_{F,Y}]^{-1}\begin{bmatrix}
        I\\
        0\\
        0
    \end{bmatrix}
\end{aligned}
\end{equation}
If automatic differentiation method is used to compute the entire framework of Neural ODEs with QP layer, then Equation \eqref{eq:qp-grad} represents the gradient of loss with respect to the variables through the QP layer.
\subsubsection{Gradients Computation of Opt-ODENet (Adjoint Method)}
The results are shown in Proposition \ref{proposition}. As to $\frac{\partial \mathcal{L}}{\partial \theta_1}$ and $\frac{\partial \mathcal{L}}{\partial \theta_2}$, the results are as follows:
\begin{equation}
\begin{aligned}
    \frac{\partial \mathcal{L}}{\partial \theta_2}&=\frac{\partial \mathcal{L}}{\partial x}\frac{\partial x}{\partial \tau(x,\theta_1,\theta_2)}\frac{\partial \tau(x,\theta_1,\theta_2)}{\partial \theta_2}\\
    &=\underbrace{\frac{\partial \mathcal{L}}{\partial x}}_{AD} \underbrace{\frac{\partial x}{\partial \tau(x,\theta_1,\theta_2)}}_{AD}\underbrace{\frac{\partial \tau(x,\theta_1,\theta_2)}{\partial h(x, \theta_2)}}_{\eqref{eq:jacobianYX}}\underbrace{\frac{\partial h(x, \theta_2)}{\partial \theta_2}}_{AD}
    \end{aligned}
\end{equation}
\begin{equation}
\begin{aligned}
    \frac{\partial \mathcal{L}}{\partial \theta_1}&=\frac{\partial \mathcal{L}}{\partial x}\frac{\partial x}{\partial \tau(x,\theta_1,\theta_2)}\frac{\partial \tau(x,\theta_1,\theta_2)}{\partial \theta_1}\\
    &=\underbrace{\frac{\partial \mathcal{L}}{\partial x}}_{AD} \underbrace{\frac{\partial x}{\partial \tau(x,\theta_1,\theta_2)}}_{AD}\underbrace{\frac{\partial \tau(x_1,\theta_1,\theta_2)}{\partial \pi(x, \theta_1)}}_{\eqref{eq:jacobianYX}}\underbrace{\frac{\partial \pi(x, \theta_1)}{\partial \theta_1}}_{AD}.
\end{aligned}
\end{equation}
\subsection{Experiments}
The source code is available at \href{https://github.com/KYMiao/L4DC-Stable-Safe-NeuralODE-Controller}{https://github.com/KYMiao/L4DC-Stable-Safe-NeuralODE-Controller}.
\subsubsection{Experiments Details of Unicycle}
\begin{figure}[t]\label{fig:hocbf-unicycle}
    \centering
    \subfigure[$\kappa = \{0.1, 0.1\}$]{
    \begin{minipage}[t]{0.5\linewidth}
        \centering
        \includegraphics[height=5.5cm]{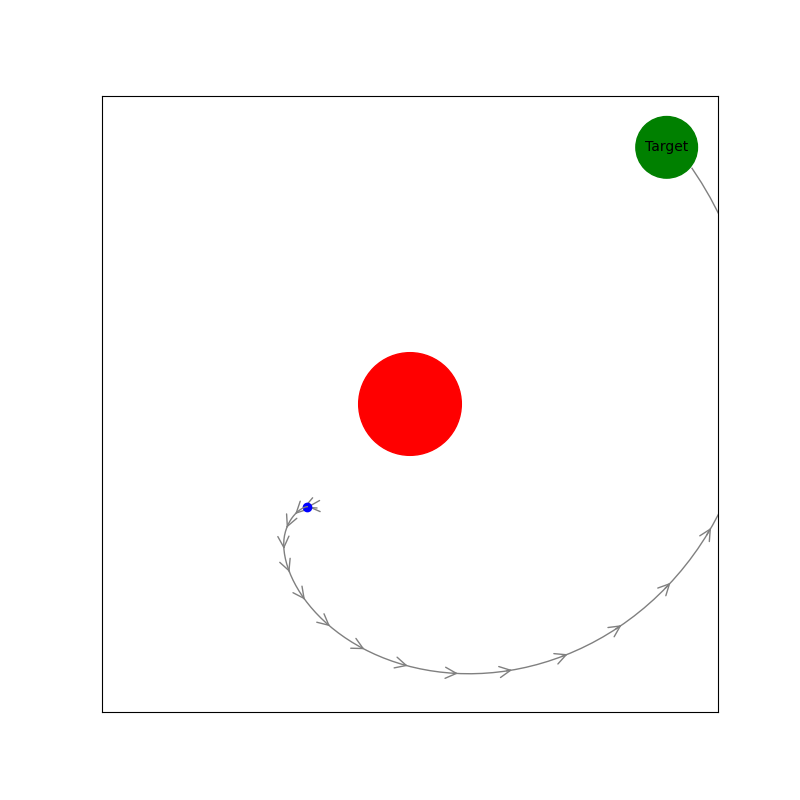}
    \end{minipage}\label{kappa0.1}}
    \hspace{-5mm}
    \subfigure[$\kappa=\{1,1\}$]{
    \begin{minipage}[t]{0.5\linewidth}
        \centering
        \includegraphics[height=5.5cm]{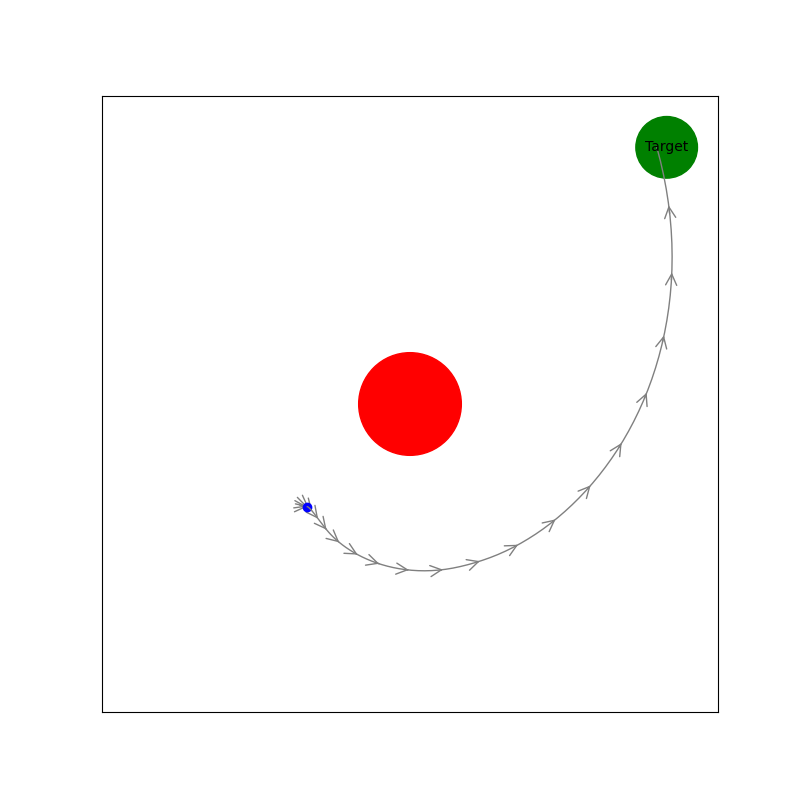}
    \end{minipage}\label{kappa1}}
    \caption{Test Trajectories under Different HOCBF-QP Settings for Unicycle}
\end{figure}

For the Unicycle environment, the controller network is designed as a two-layer neural network with a hidden state dimension of 64. The training is performed with a batch size of 32 over 100 epochs. The Neural ODE is solved using the Euler method, with a time interval of $[0, 1]$ and a step size of 0.01.\\
The setting in Equation \eqref{eq:unicycle-modify} is also set up to deal with the different relative degrees for \eqref{eq:unicycle-dynamics}. Another way is modify the dynamics as follows: 
\begin{equation}
    \left[ \begin{array}{c}
                \dot x \\
                \dot y\\
                \dot \theta\\
                \dot v
        \end{array}\right] = \left[ \begin{array}{c}
                v \cos(\theta) \\
                v \sin(\theta\\
                0\\
                0
        \end{array}\right] + \left[ \begin{array}{cc}
        0  & 0 \\
        0 & 0 \\
        1.0 & 0\\
        0 & 1.0 
        \end{array}
        \right ] \left[ \begin{array}{c}
                v \\
                \omega
        \end{array}\right]
\end{equation}
In this case, HOCBF should be used. We show some results to illustrate the importance of the choice of class-$\mathcal{K}$ function in Figure \ref{fig:hocbf-unicycle}.
In this case, a larger $\kappa$ typically delays the system’s response to CBF constraints, giving priority to the objective. This often results in the system adhering more closely to obstacles, as it adjusts later to safety constraints.
\subsubsection{Experiments Details of Simulated Cars}
\begin{figure}
    \centering
    \includegraphics[width=1.0\linewidth]{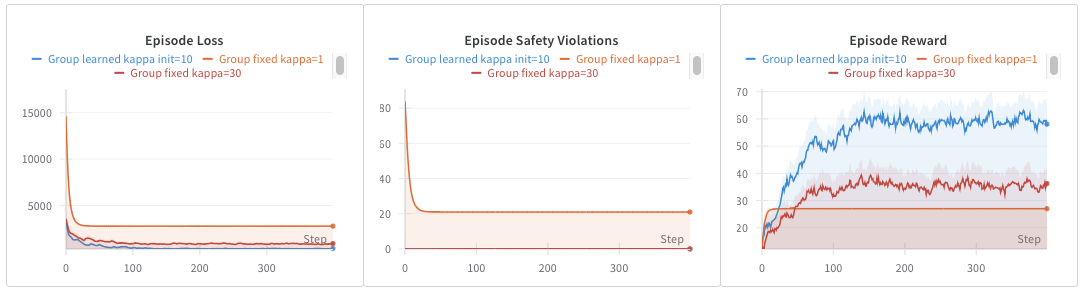}
    \caption{Training results on Simulated Cars under Different HOCBF-QP Settings}
    \label{fig:hocbf-compare}
\end{figure}

This environment, simulating a chain of five cars following each other on a straight road. The objective is to control the acceleration of the $4^{th}$ car to maintain a desired distance from the $3^{rd}$ car while avoiding collisions with other cars. The real dynamics of all cars except the $4^{th}$ one is given by:
\begin{equation*}
    \begin{array}{l}
        \dot{x}_i(t)=
        \left[ \begin{array}{c}
        v_i(t) \\
        0
        \end{array}
        \right ] +
        \left[ \begin{array}{c}
            0  \\
            1+d_i 
            \end{array}
            \right ] a_i(t)   \qquad  \forall i \in \{1,2,3,5\}.
    \end{array}
\end{equation*}
Each state of the system is denoted as $x_i(t) = [p_i(t),v_i(t)]^{\top}$, indicating the position $p_i(t)$ and velocity $v_i(t)$ of the $i^{th}$ car at the time $t$, $d_i=0.1$. The time interval used in this experiment is $0.02$s. The predefined velocity of the $1^{st}$ car is $v_{s} - 4\sin (t)$, where $v_{s}=3.0$. Its acceleration is given as $a_1(t)=k_v(v_{s} - 4\sin (t) -v_1(t))$ where $k_v=4.0$. Accelerations of Car 2 and 3 are given by:
    \begin{equation*}
    a_i(t)\!=\!\left\{
\begin{aligned}  
&\!k_v(v_{s}\!-\!v_i(t))\!-\!k_b(p_{i-1}(t)\!-\!p_i(t))\,\,if\,|p_{i-1}(t)\!-\!p_i(t)|\! <\! 6.5\\
&\!k_v(v_{s}\!-\!v_i(t))\,\,\,\,\,\,\,\,\,\,\,\,\,\,\,\,\,\,\,\,\,\,\,\,\,\,\,\,\,\,\,\,\,\,\,\,\,\,\,\,\,\,\,\,\,otherwise, \\
\end{aligned}
\right.
\end{equation*}    
where $k_b=20.0$ and $i=2,3$. The acceleration of the $5^{th}$ car is:
\begin{equation*}
    a_5(t)\!=\!\left\{
\begin{aligned}
&\!k_v(v_{s}\!-\!v_5(t))\!-\!k_b(p_3(t)\!-\!p_5(t))\,\,if\,|p_3(t)-p_5(t)| \!<\! 13.0\\
&\!k_v(v_{s}\!-\!v_5(t))\,\,\,\,\,\,\,\,\,\,\,\,\,\,\,\,\,\,\,\,\,\,\,\,\,\,\,\,\,\,\,\,\,\,\,\,\,\,\,\,\,\,\,\,\,otherwise. \\
\end{aligned}
\right.
\end{equation*}
The model of the $4^{th}$ car is as follows: 

        \begin{equation*}
         \dot{x}_4(t)=
         \left[ \begin{array}{c}
         v_4(t) \\
         0
         \end{array}
         \right ] +
         \left[ \begin{array}{c}
         0 \\
         1.0
         \end{array}
         \right ] u(t) ,
         \end{equation*} 
where $u(t))$ is the acceleration of the $4^{th}$ car, and also the control signal generated by the controller at the time $t$.\\
$d(t) =p_3(t)-p_4(t)$ represents the distance between the $3^{rd}$ and $4^{th}$ car, and the objective is to let $d(t)$ fall within $[5.5,6.5]$ as a desired region. Thus, the CLF is determined as $\left\lVert d(t) - d_{\text{desired}}\right\rVert $, where $d_{\text{desired}}=6$. CBFs are defined as $h_1(x(t)) =p_3(t)-p_4(t) - \delta$ and $h_2(x(t)) =p_4(t)-p_5(t) - \delta$, with $\delta$ being the minimum required distance between the cars.\\
The controller network is designed as a four-layer neural network with hidden state dimensions of 128, 64, 32 and 12. The training is performed over 400 epochs. The Neural ODE is solved using the Euler method, with a time interval of $[0, 4]$ and a step size of 0.02.\\
The training results under different HOCBF-QP settings are shown in Figure \ref{fig:hocbf-compare} where the reward represents how often $d(t)$ falls with desired region.
It can be found that when $\kappa = \{2,1\}$, the whole training seems meaningless, the reason may be that the CBF cannot work normally and the car stucks when $\kappa$ is too small and therefore contradicts with the whole setting, and CLF cannot help the car get out of this circumstance (the fourth car cannot monotonically going further from both the third and the fifth car at the same time and therefore certain infeasibility happens, which makes the whole training problematic. - however in this case we can consider learning different $\kappa$ for different safety constraints for improvement) When $\kappa=\{60, 900\}$, the safety constraints are valid and the safety violations are 0. For the reward, this case works worse than the case that $\kappa$ is learned as $\{18.6, 96.04\}$. Hence, the effect of $\kappa$ is not as straightforward as `larger $\kappa$ equals more aggressive behavior.' While a larger $\kappa$ often delays the system’s response to CBF constraints, allowing it to prioritize the CLF objective and potentially `stick closer' to obstacles, this does not universally translate to faster or more efficient target convergence. The interaction between CBF and CLF objectives is context-dependent. For example, when the target lies near an obstacle, a larger $\kappa$ might help the system avoid detours by adhering closely to the safety boundary. However, if the target direction naturally steers the system away from the obstacle, increasing $\kappa$ adds little value and could even result in suboptimal trajectories due to delayed responses to other constraints. Thus, the impact of class-$\mathcal{K}$ function is dynamic and contextual, and in some cases it is important to learn it rather than fix it to better accommodate different CBF-CLF interactions, especially for hard constraints.
\end{document}